\documentclass[10pt,journal]{IEEEtran}%

\usepackage[T1]{fontenc}

\ifCLASSINFOpdf
\else
\fi
\usepackage{amsmath}
\usepackage{courier}
\usepackage{graphicx}
\usepackage[T1]{fontenc} 
\usepackage{amsmath}
\usepackage{mathrsfs}
\usepackage{bm} 
\usepackage{graphicx}
\usepackage{algorithm}
\usepackage{algorithmicx}
\usepackage{algpseudocode}
\usepackage{array}
\usepackage{tabularx}
\usepackage{balance}
\usepackage{makecell}
\usepackage{color}
\usepackage{multirow}
\usepackage{tabularx}
\usepackage{amsfonts}
\usepackage{mathrsfs}
\usepackage{cite}
\usepackage{color}
\usepackage{subcaption}

\usepackage{xpatch}
\makeatletter
\newcommand{\multiline}[1]{%
  \begin{tabularx}{\dimexpr\linewidth-\ALG@thistlm}[t]{@{}X@{}}
    #1
  \end{tabularx}
}
\makeatother

\algnewcommand{\LeftComment}[1]{\Statex \(\triangleright\) #1}

\DeclareMathOperator*{\argmax}{arg\,max}

\begin{document}
%
\title{Online Service Migration in Edge Computing with Incomplete Information: A Deep Recurrent Actor-Critic Method}
%
%
%

\author{Jin~Wang,
        Jia~Hu,
        and~Geyong~Min
\thanks{College of Engineering Mathematics and Physical Sciences, University of Exeter, UK}
\thanks{Email:\{jw855, j.hu, g.min\}@exeter.ac.uk }
}

\makeatletter
\long\def\@IEEEtitleabstractindextextbox#1{\parbox{0.922\textwidth}{#1}} 
\makeatother

\IEEEtitleabstractindextext{%
\begin{abstract}
Multi-access Edge Computing (MEC) is an emerging computing paradigm that extends cloud computing to the network edge (e.g., base stations, MEC servers) to support  resource-intensive applications on mobile devices. As a crucial problem in MEC, service migration needs to decide where to migrate user services for maintaining high Quality-of-Service (QoS), when users roam between MEC servers with limited coverage and capacity. However, finding an optimal migration policy is intractable due to the highly dynamic MEC environment and user mobility. Many existing works make centralized migration decisions based on complete system-level information, which can be time-consuming and suffer from the scalability issue with the rapidly increasing number of mobile users. To address these challenges, we propose a new learning-driven method, namely Deep Recurrent Actor-Critic based service Migration (DRACM), which is user-centric and can make effective online migration decisions given incomplete system-level information. Specifically, the service migration problem is modeled as a Partially Observable Markov Decision Process (POMDP). To solve the POMDP, we design an encoder network that combines a Long Short-Term Memory (LSTM) and an embedding matrix for effective extraction of hidden information. We then propose a tailored off-policy actor-critic algorithm with a clipped surrogate objective for efficient training. Results from extensive experiments based on real-world mobility traces demonstrate that our method consistently outperforms both the heuristic and state-of-the-art learning-driven algorithms, and achieves near-optimal results on various MEC scenarios. 
\end{abstract}
}
\maketitle
\IEEEdisplaynontitleabstractindextext

\IEEEpeerreviewmaketitle

\section{Introduction}

\IEEEPARstart{R}{ecent} years have witnessed a booming of emerging mobile applications such as augmented reality, virtual reality, and interactive gaming. These types of applications require intensive computing power for real-time processing, which often exceeds the limited computing and storage capabilities of mobile devices. To resolve this issue, Multi-access Edge Computing (MEC) \cite{sabella2019developing}, a new computing paradigm, was proposed to meet the ever-increasing demands for the Quality-of-Service (QoS) of mobile applications. MEC provides many computing and storage resources at the network edge (close to users), which can effectively cut down the application latency and improve the QoS. Specifically, a mobile application empowered by the MEC consists of a front-end component running on mobile devices, and a back-end service that runs the tasks offloaded from the application on MEC servers \cite{rejiba2019survey}. In this way, the MEC enables mobile devices with limited processing power to run complex applications with satisfied QoS. 

When considering the user mobility along with the limited coverage of MEC servers, the communications between a mobile user and the user service running on an edge server may go through multiple hops, which would severely affect the QoS. To address this problem, the service could be dynamically migrated to a more suitable MEC server so that the QoS is maintained. Unfortunately, finding an optimal migration policy for such a problem is non-trivial, due to the complex system dynamics and user mobility. Many existing works \cite{ouyang2018follow,wang2019delay,wu2020mobility,wang2019dynamic,ning2020distributed} proposed service migration solutions based on Markov Decision Process (MDP) or Lyapunov optimization under the assumption of knowing the complete system-level information (e.g., available computation resources of MEC servers, profiles of offloaded tasks, and backhaul network conditions). Thus, they designed centralized controllers (i.e., controllers are placed on edge servers or central cloud) that make migration decisions for mobile users in the MEC system. 

The aforementioned methods have two potential drawbacks: 1) in a real-world MEC system, gathering complete system-level information can be difficult and time-consuming; 2) the centralized control approach will have the scalability issue since its time complexity rapidly increases with the number of mobile users. To address the above issues, some works proposed decentralized service migration methods based on contextual Multi-Armed Bandit (MAB) \cite{sun2017emm,ouyang2019adaptive,sun2018learning}, where the migration decisions are made by the user side with partially observed information. However, they did not consider the intrinsically large state space and complex dynamics in the MEC system, which may lead to unsatisfactory performances. A recent work \cite{yuan2020joint} modeled the joint optimization problem of service migration and path selection as a partially observable Markov decision process (POMDP) solved by independent Q-learning, which can be unstable and inefficient when handling the MEC environment with continuous state space (e.g., data size, CPU cycle, workload) and complex system dynamics. 

To address the above challenges, we propose a Deep Recurrent Actor-Critic based service Migration (DRACM) method,  which is user-centric and can learn to make online migration decisions with incomplete system-level information based on Deep Reinforcement Learning (DRL). DRL is able to solve complex decision-making problems in various areas, including robotics \cite{gu2017deep}, games \cite{ye2020mastering}, networks \cite{chinchali2018cellular}, etc., making it an attractive approach. Distinguished from the existing works, we model the service migration problem as a POMDP with continuous state space and develop a tailored off-policy actor-critic algorithm to efficiently solve the POMDP. The main contributions of this work are listed as follows:

\begin{itemize}
 \item We model the service migration problem as a POMDP to capture the intrinsically complex system dynamics in the MEC. We solve the POMDP by proposing a novel off-policy actor-critic method, DRACM. Specifically, our method is model-free and can quickly learn effective migration policies through end-to-end reinforcement learning (RL), where the agent makes online migration decisions based on the sampled raw data from the MEC environment with minimal human expertise.  
 \item A new encoder network that combines a Long Short-Term Memory (LSTM) and an embedding matrix is designed to effectively extract the hidden information from the sampled histories. Moreover, a tailored off-policy actor-critic algorithm with a clipped surrogate objective function is developed to substantially stabilize the training and improve the performance.
 \item We demonstrate how to implement the DRACM in an emerging MEC framework, where the migration decisions can be made online through the inference of the policy network, while the training of the policy network can be offline, saving the cost of directly interacting with the MEC environment.    
 \item Extensive experiments are conducted to evaluate the performance of the DRACM using real-world mobility traces. The results demonstrate that the DRACM has a stable training process and high adaptivity to different scenarios, while outperforms the online baseline algorithms, and achieves near-optimal results.
\end{itemize}

The remainder of this paper is organized as follows. Section \ref{sec::problem_form} gives the problem formulation of service migration. Section \ref{sec::method} presents the DRL backgrounds, POMDP modeling for service migration, details of the DRACM algorithm, and the implementation of the DRACM in the emerging MEC system. In Section \ref{sec::experiments}, we evaluate the performance of the DRACM and five baseline algorithms on two real-world mobility traces with various MEC scenarios. We then review the related works in Section \ref{sec::related_work}. Finally, Section \ref{sec::conclusion} draws conclusions.

\section{Problem Formulation of Service Migration}
\label{sec::problem_form}
\begin{figure}[t]
    \centering
    \includegraphics[width=2.85in]{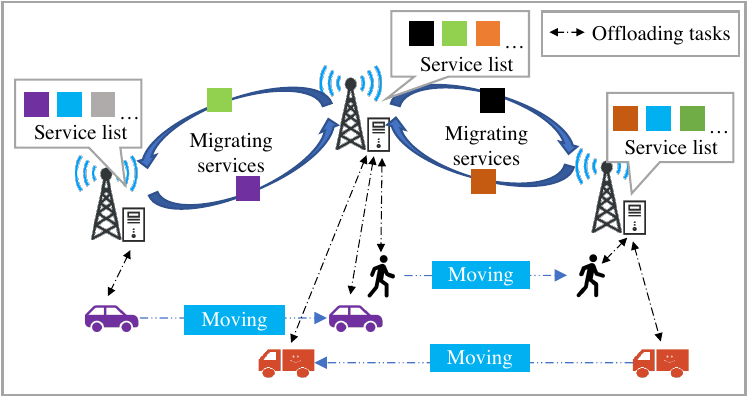}
    \centering\caption{An example of service migration in MEC.}
    \label{migration_example}
\end{figure}

As shown in Fig. \ref{migration_example}, we consider a typical scenario where mobile users move in a geographical area covered by a set of MEC servers, $\mathcal{M}$, each of which is co-located with a base station. In the MEC system, mobile users can offload their computation tasks to the services provided by MEC servers. We define the MEC server that runs the service of a mobile user as the user's \textit{serving node}, and the MEC server that directly connects with the mobile user as the user's \textit{local server}. In general, the MEC servers are interconnected via stable backhaul links, thus the mobile user can still access its service via multi-hop communication among MEC servers when it is no longer directly connected to the serving node. To maintain satisfactory QoS, the service should be dynamically migrated among the MEC servers as the user moves. In this paper, we use latency as the measurement for the QoS that consists of migration, computation, and communication delays.

We consider a time-slotted model, where a user's location may only change at the beginning of each time slot. The time-slotted model is widely used to address the service migration problem \cite{wang2019dynamic,ouyang2019adaptive,wang2019delay}, which can be regarded as a sampled version of a continuous-time model. When a mobile user changes location, the user makes the migration decision for the current service and then offloads computation tasks to the serving node for processing. Denote the migration decision at time slot $t$ as $a_t$ ($a_t \in \mathcal{M}$), where $a_t$ can be any of the MEC servers in this area. In general, the migration, computation, and communication delays are expressed as follows. 

\textbf{Migration delay: } The migration delay is incurred when a service is moved out from the previous serving node. In general, the migration delay $B(d_t) = m_t^c d_t$ is a non-decreasing function of $d_t$ \cite{wang2019dynamic,ouyang2019adaptive,wang2016dynamic}, where $d_t$ is the hop distance between the current serving node $a_t$ and the previous one $a_{t-1}$, and $m_t^c$ is the coefficient of migration delay. The migration delay can capture the service interruption time during migration, which increases with the hop distance due to the involved propagation and switching delay of service data transmission. 

\textbf{Computation delay: }At each time slot, the mobile user may offload computation tasks to the serving node for processing. The computing resources of MEC servers are shared by multiple mobile users to process their applications. At time slot $t$, we denote the sum of the required CPU cycles for processing the offloaded tasks as $c_t$, the workload of the serving node as $w_{t}^{a_t}$, and the total computing capacity of the serving node as $f^{a_t}$. We consider a weighted resource allocation strategy on each MEC server, where tasks are allocated with computation resources proportional to their required CPU cycles. Therefore, the computation delay of running the offloaded tasks at time slot $t$, can be calculated as 
\begin{equation}
\label{compute_cost}
 D(a_t) = \frac{c_t}{ ( \frac{c_t}{w_{t}^{a_t} + c_t} f^{a_t})} = \frac{w_{t}^{a_t} + c_t}{f^{a_t}}.  
\end{equation}

\textbf{Communication delay: } After migrating the service, the communication delay is incurred when the mobile user offloads computation tasks to the serving node. Generally, the communication delay consists of two parts: access delay between the mobile user and the local server, and backhaul delay between the local server and the serving node. The access delay is determined by the wireless environment and the data size of the offloaded tasks. At time slot $t$, we denote the data size of the offloaded tasks as $data_t$, the average upload rate of the wireless channel as $\rho_t$. Hence, the access delay can be expressed as
\begin{equation}
R(data_t) = \frac{data_t }{\rho_t}.
\end{equation}
 While the backhaul delay is incurred by data transmission, propagation, processing, and queuing between the serving node and the local server through backhaul networks, which mainly depends on the hop distance along the shortest communication path and the data size of the offloaded tasks \cite{yuan2020joint,wang2019dynamic,ouyang2019adaptive}. We denote the local server at time slot $t$ as $u_t$ ($u_t \in M$) and the hop distance between the serving node $a_t$ and the local server $u_t$ as $y_t$. The bandwidth of the outgoing link of the local server is denoted as $\eta_t$. Generally, the transmission delay of the computation results can be ignored because of the small data size. Consequently, the backhaul delay can be given by
\begin{equation}
P(y_t, data_t)=\left\{
\begin{aligned}
&0, \ \ \  & {\rm if} \ \ \  y_t = 0,  \\
 &\frac{data_t}{\eta_t} + 2\lambda_{\rm bh} y_t,  \ \ \  & {\rm if} \ \ \  y_t \ne 0, \\
\end{aligned}
\right.
\end{equation}
where $\lambda_{\rm bh}$ is a coefficient of the backhaul delay \cite{yuan2020joint}. Especially, when the serving node and mobile user are directly connected ($y_t = 0$), there is no backhaul cost. Overall, the total communication delay at time slot $t$ can be obtained by 
\begin{equation}
E(y_t, data_t) = R(data_t) + P(y_t, data_t).
\end{equation}

Given a finite time horizon $T$, our objective for the service migration problem is to obtain optimal migration decisions, $\{a_1, a_2, ..., a_T\}$, so that the sum of all the above costs (i.e., total latency) is minimal. Formally, the objective is expressed as:
\begin{equation}
\begin{split}
  & \min_{a_0, a_1, ..., a_T} \sum_{t=0}^{T} B(d_t) + D(a_t) + E(y_t, data_t), \\
    & {\rm s.t.} \ \ \ a_t \in \mathcal{M}.
\end{split}
\label{objective}
\end{equation}

Obtaining the optimal solution for the above objective is challenging, which requires user mobility and complete system-level information over the entire time horizon. However, in real-world scenarios, it is impractical to gather all the relative information in advance. To address this challenge, we propose a learning-based online service migration method that can make efficient migration decisions based on partially observed information. In the next section, we present our solution in detail. 

\section{Online Service Migration with Incomplete Information}
\label{sec::method}
Service migration in MEC is intrinsically a sequential decision-making problem with a partially observable environment (i.e., with incomplete system information), which can be naturally modeled as a POMDP. We solve the POMDP with the proposed DRACM method to provide effective online migration decisions. Before presenting the details of our solution, we first introduce the necessary backgrounds.

\begin{figure}[t]
    \centering
    \includegraphics[width=2.0in]{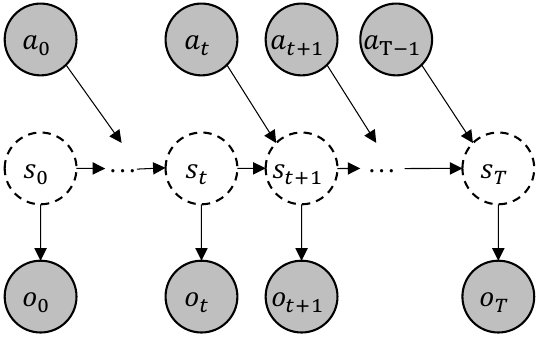}
    \centering\caption{Graphical model of POMDP.}
    \label{pomdp_model}
\end{figure}

\subsection{Backgrounds of RL and POMDP}

\textbf{Reinforcement learning: }RL can solve sequential decision-making problems by learning from interaction with the environment. In general, RL uses the formal framework of MDP, which is defined by a tuple $(\mathcal{S}, \mathcal{A}, \mathcal{P}, \mathcal{R}, \gamma)$, to represent the interaction between a learning agent and its environment. Specifically, $\mathcal{S}$ is the state space, $\mathcal{A}$ denotes the action space, $\mathcal{P}$ is the environment dynamics,  $\mathcal{R}$ represents the reward function, and $\gamma$ is the discounted factor. The policy, $\pi(\cdot|s_t)$, represents the distribution over actions given a state $s_t$. The return from state $s_t$, which is defined as  $G_t(\tau) = \sum_{i=t}^{T} \gamma^{i-t} r_t$, is the sum of discounted rewards along a trajectory $\tau := \{s_0, a_0, r_0, s_1, a_1, r_1, ..., s_T, a_T, r_T\}$. The goal of RL is to find an optimal policy $\pi^*$, so that the expected return, $\mathbb{E}_{\tau \sim p(\tau|\pi^*)}[G_0(\tau)]$, is maximal. 

The action-value function is defined by the expected return after taking an action $a_t$ in state $s_t$ and thereafter following policy $\pi$, $q_{\pi}(s_t, a_t) = \mathbb{E}_{\pi} \left [ G_{t} | s_t, a_t \right ]$. An optimal action-value function, which is defined as $q^*(s_t, a_t) = \max_{\pi} q_{\pi}(s_t, a_t)$, is the maximum action value achievable by any policy for state $s_t$ and action $a_t$. The valued-based DRL methods (e.g., deep Q-learning (DQL) \cite{mnih2015human}) use the deep neural network to approximate the optimal action-value function, $q^*(s_t, a_t; \theta^Q)$ where $\theta^Q$ are parameters of the deep neural network. They obtain the optimal policy by greedily selecting the action with maximal action value, where $a_t = \argmax_{a} q^*(s_t, a; \theta^Q)$. {\color{black} However, since DQL indirectly obtains a deterministic policy by training the Q-network (i.e., a neural network that is used approximate the action-value function), it generally has a low convergence rate \cite{mnih2016asynchronous}. The complex state space and large action space of the MEC environment exacerbate this issue. Besides, the training target of DQL is obtained by one-step bootstrapping of the Q-network, which can be a highly biased estimation of the true action values. Introducing bias may harm the convergence of the algorithm, or cause converging to sub-optimal solutions. The above issues make DQL unfit to solve the service migration problem since the learned migration policies may lead to unsatisfied performance. In contrast, the policy-based methods (e.g., asynchronous actor-critic \cite{mnih2016asynchronous} and proximal policy optimization \cite{schulman2017proximal}) provide good convergence property for dealing with the complex state and action space of the environment. They directly parameterized the stochastic policy with a deep neural network rather than using deterministic policy derived from the action-value function. The parameters of the policy network are updated by performing gradient ascent on $\mathbb{E}[G_0(\tau)]$. In this paper, we build our method (i.e., the DRACM) based on the policy-based methods and show the performance comparison between the DQL-based method and the DRACM in Section \ref{sec::experiments}. } 

\textbf{Partially Observable Markov Decision Process: }MDP assumes that states include complete information for decision-making. However, in many real-world scenarios, observing such states is intractable. Therefore, the POMDP, an extension of MDP, is proposed as a general model for the sequential decision-making problem with a partially observable environment, which is defined by a tuple $(\mathcal{O}, \mathcal{S}, \mathcal{A}, \mathcal{P}, \mathcal{R}, \mathcal{U}, \gamma)$. Fig. \ref{pomdp_model} shows the graphical model of POMDP. Specifically, the state $s_t \in \mathcal{S}$ is latent and the observation $o_t \in \mathcal{O}$ contains partial information of the latent state $s_t$. $\mathcal{U}(o_t | a_{t-1}, s_t)$ represents the observation distribution, which gives the probability of observing $o_t$ if action $a_{t-1}$ is performed and the resulting state is $s_t$. Since the state is latent, the learning agent cannot choose its action directly based on the state. Alternatively, it has to consider a complete history of its past actions and observations to choose its current action. Specifically, the history up to time step $t$ is defined by $H_t = \{o_0, a_0, ..., o_{t-1}, a_{t-1}, o_t\}$. Therefore, the key for RL-based methods to solve the POMDP is how to effectively infer the latent state based on the history, which is defined by $p(s_t| o_{\le t}, a_{<t})$. In the literature, some RL methods \cite{hausknecht2015deep,zhu2018improving} assume the latent states as deterministic states, which encode the whole history by RNN and use the hidden state of RNN as input to the policy. Other works \cite{watter2015embed,igl2018deep,zhang2019solar} explicitly infer the belief state that is defined by the distribution over latent states (i.e., stochastic latent state)  given the history and sampling latent state from the distribution as input to the policy. We use LSTM for latent information extraction, which can achieve excellent performance and is much easier to be implemented in MEC scenarios compared to methods based on inferring the belief state. In the next subsection, we present the motivations of POMDP modeling for service migration problem and the detailed definition of the model. 

\subsection{POMDP modeling for service migration problem}
Key factors that affect the migration decision of a mobile user at a time slot are the mobility of the user, the offloading tasks' profile, the workloads of edge servers, and the resource allocations of edge servers, etc. Ideally, the user can make optimal migration decisions if knowing complete information related to the decision-making process. However, some information are hard to obtain for the user side. For example, at each time slot, the workloads of edge servers are determined by the task requests from their associated mobile users and the available computation resources of edge servers. However, it is unlikely for a mobile user to get such information. To make effective decisions based on partially observable information, POMDP is a natural choice to model the problem, which gives the agent the ability to effectively estimate the outcome of its actions even when it cannot exactly observe the state of its environment. In our POMDP modeling, the mobile user treats the unobserved information (e.g., workloads and resource allocations of MEC servers) as a part of the latent state. Differing from the simplified model such as MAB, POMDP does not ignore the intrinsic large state space and complex dynamics of the service migration problem, thus solving the POMDP can result in more effective decisions.    

The detailed POMDP model of service migration is defined as follows:
\begin{itemize}
\item \textit{Observation}: The observation contains information that is accessible from the user side, which is defined by a tuple of the local server $u_t$, the transmission rate of wireless network $\rho_t$, the required CPU cycles of computation tasks $c_t$, and the sizes of transmission data, $data_t$: 
	\begin{equation}
		o_t := (u_t, \rho_t, c_t, data_t).
		\label{obs_definition}
	\end{equation}
	Note that the geographical location of the mobile user is an indirect factor that affects the migration decisions, which determines the local server associated with the mobile user and affects the transmission rate (included in our definition of the observation, Eq. (\ref{obs_definition})). Therefore, we define the local server $u_t$ as a component of the observation rather than the geographical location of the mobile user.
\item \textit{Action}: At each time slot, the service can be migrated to any of the MEC servers in the area. Therefore, an action is defined as $a_t \in \mathcal{M}$. 
\item \textit{Reward}: The reward at each time slot is defined as the negative sum of migration, computation, and communication delays, which is formally expressed as 
\begin{equation}
r_t := -\left (B(d_t) + D(a_t) + E(y_t, data_t) \right).
\end{equation}
\end{itemize}

Solving the above POMDP is non-trivial due to the complex dynamics and continuous state space of the MEC environment. In the next subsection, we present our method, DRACM, to solve the above POMDP.  

\subsection{Deep Recurrent Actor-Critic based service Migration (DRACM)}
\begin{figure*}[t]
    \centering
    \includegraphics[width=5.8in]{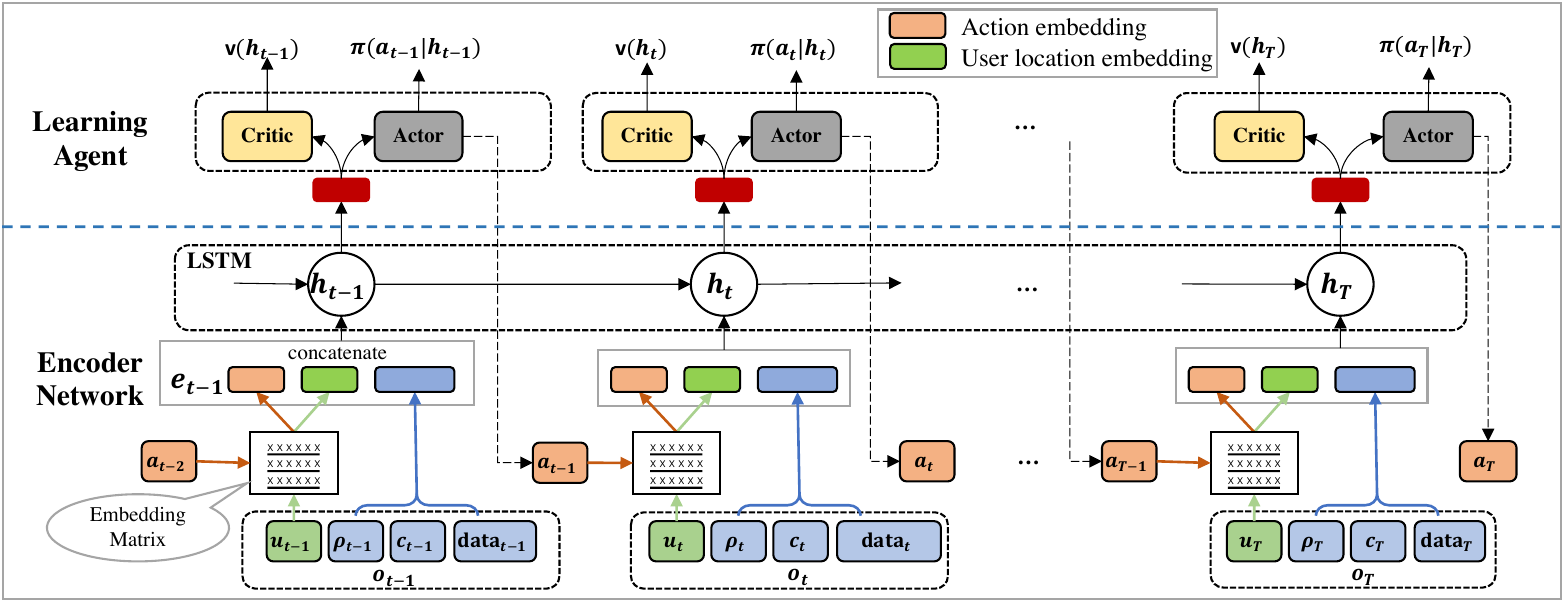}
    \centering\caption{The architecture of the DRACM.}
    \label{network_archi}
\end{figure*}

Fig. \ref{network_archi} shows the overall architecture of the DRACM, which follows an end-to-end principle with raw history sampled from the environment as input and the migration decisions as output. The DRACM consists of two parts: the encoder network and the learning agent, where the encoder network learns to effectively represent the latent state of the POMDP based on the history and the learning agent learns to make effective migration decisions. {\color{black} The goal of the encoder network is to infer the latent state of the POMDP based on the observed history:
\begin{equation}
p(s_{1:T}|o_{1:T}, a_{1:T-1}) = \prod_{t=1}^T p(s_t| s_{t-1}, a_{t-1}, o_t)
\end{equation}
Here, we include a LSTM to approximate the above function where the hidden state of the LSTM, $h_t$, is used to represent the latent state $s_t$ of the POMDP, thus we have
\begin{equation}
    \label{lstm_func}
    h_t = f_{\rm enc} ([o_{\le t}, a_{< t}]; \theta)= f_{\rm enc} ([o_t, a_{t-1}], h_{t-1}; \theta),
\end{equation}
where $t \in [1, T]$, $f_{\rm enc}$ and $\theta$ represent the inner process and parameters of the encoder network, respectively. }

To improve the representation ability of the features $u_t$ and $a_{t-1}$, we convert them into embeddings by looking up a trainable $|\mathcal{M}| \times d_e$ matrix, where $d_e$ is the dimension of embedding vectors. Subsequently, the action embedding, user location embedding, and the rest components of the observation are concatenated as a vector, $e_t$, feeding into the LSTM to produce the hidden state $h_t$.

The learning agent is based on a standard actor-critic structure. Both actor and critic are parametrized by neural networks with the hidden state $h_t$ as input. We denote $\phi$ and $\psi$ as the parameters of actor and critic networks, respectively. The actor network aims at approximating the policy, $\pi(a_t|h_t; \phi)$, which outputs a distribution over the action space at time step $t$ given $h_t$. Meanwhile, the critic network, $v(h_t, \psi)$, approximates the value function that is an estimation of the expected return when starting in $h_t$ and following the policy $\pi$ thereafter.  

Denote the trajectory sampled from the environment following policy $\pi$ as $\tau = \{o_0, a_0, r_0, ..., o_T, a_T, r_T\}$. The critic network can be updated by minimizing the mean square error of one-step temporal differences $\delta_t$ based on the sampled trajectories, which is formally defined as
\begin{equation}
    L^{\rm critic}(\psi, \theta) = \mathbb{E}_{\tau \sim p(\tau|\pi)} \left [ \sum_{t=0}^T \delta_t^2 \right]
    \label{critic_network_obj}, 
\end{equation}
\begin{equation}
\delta_t =  r+\gamma v(h_{t+1};\psi) - v(h_t;\psi),
\label{critic_network_obj_2}
\end{equation}
where the $h_t$ can be obtained by Eq. (\ref{lstm_func}). The objective of the actor is to find an optimal policy that maximizes the accumulated reward, which can be formally expressed as 
\begin{equation}
L^{\rm act}(\phi, \theta) = \mathbb{E}_{\tau \sim p(\tau|\pi)} \left [ \sum_{t=0}^T \gamma^t r_t\right].
\label{actor_objective}
\end{equation}
The optimal policy can then be obtained by gradient assent through policy gradient with one-step actor-critic \cite{sutton2018reinforcement}, where the gradient of the above objective function can be calculated by
\begin{equation}
\begin{split}
& \nabla_{\theta, \phi}L^{\rm act} = \mathbb{E}_{\tau \sim p(\tau|\pi)} \left [ \sum_{t=0}^{T}\delta_t \nabla_{\theta, \phi}\log \pi(a_t|h_t; \phi) \right ]. \\
\label{pg_target}
\end{split}
\end{equation}

\begin{algorithm}[t]
  \caption{Deep Recurrent Actor-Critic based service Migration (DRACM)}
  \label{DRACM}
  Initialize the parameters of behavior policy $\phi'$, behavior encoder network $\theta'$, target policy $\phi$, target encoder network  $\theta$, and critic network $\psi$, 
  \begin{algorithmic}[1]
    \For { $k = 0, 1, 2, ..., n $}
    \LeftComment{ \% Start sampling process \% }
    \State \multiline{Synchronize the parameters: $\theta' \leftarrow \theta$, $\phi' \leftarrow \phi$.}
    \State \multiline{Sample a set of trajectories $D_{\tau} = \{\tau_0, \tau_1, ... \tau_n\}$ by running the behavior policy $\pi'(a_t|h'_t; \phi')$ in the environment, where $h'_t = f_{\rm enc} ([o_{\le t}, a_{< t}]; \theta')$.}
    \State \multiline{Compute the advantage estimator, $\hat{A}_t$, according to Eq. (\ref{pg_adv}). }

    \LeftComment{ \% Start target policy updating process \% }
    \For {$j = 0, 1, 2, ..., m$}
    \State \multiline{Update the parameters of encoder network $\theta$, target policy network $\phi$, and critic network $\psi$, \\
        $\ \ \ \theta \leftarrow \theta + \nabla_\theta L^{\rm act}_{\rm c}(\phi, \theta) - \nabla_\theta L^{\rm critic}(\psi, \theta)$, \\ 
        $\ \ \ \phi \leftarrow \phi + \nabla_\phi L^{\rm act}_{\rm c}(\phi, \theta)$, \\
        $\ \ \ \psi \leftarrow \psi - \nabla_\psi L^{\rm critic}(\psi, \theta)$, \\
    by mini-batch gradient updates  based on collected trajectories $D_{\tau}$ with \textit{Adam}.}
    \EndFor
    \EndFor
  \end{algorithmic}
\end{algorithm}

However, directly applying the above on-policy (i.e., using the same policy for training and sampling) objective has some drawbacks when solving the service migration problem. First, we cannot train the policy network offline with mini-batches by using on-policy objective. This can lead to severe sample efficiency problem, since the learning agent needs to resample trajectories from the environment after each gradient update. Especially, in the MEC system, frequently interacting with the environment to get the training samples is costly. Second, the on-policy objective has limited exploring ability, thus the policy can easily get stuck in a local optima. {\color{black} Third, to reduce the variance of the objective function, Eq. (\ref{pg_target}) includes a biased estimator $\delta_t$. However, introducing bias may harm the convergence of the algorithm.} To address the above problems, we design an off-policy (i.e., training a policy different from that was used to sample the data) algorithm that can train the policy with mini-batches and reduce the interaction frequency with the environment. Inspired by the previous works on RL \cite{schulman2017proximal,haarnoja2018soft,schulman2016high}, we introduce an off-policy training method  with a surrogate objective as follows:  
\begin{equation}
\label{surrogate_obj}
L^{\rm act}_{\rm c}(\phi, \theta) = \mathbb{E}_{\tau \sim p(\tau|\pi')} \left [\sum_{t=0}^T g_{\rm clip}(\pi'_t, \pi_t, \hat{A}_t) + c_{h} \mathcal{H}(\pi_t)  \right], 
\end{equation}

\begin{equation}
 g_{\rm clip}(\pi'_t, \pi_t, \hat{A}_t) = \min \left (\frac{\pi_t}{\pi_t'}\hat{A}_t, {\rm clip}_{1-\epsilon}^{1+\epsilon} \left (\frac{\pi_t}{\pi_t'} \right) \hat{A}_t \right ),
 \label{clip_grad}
\end{equation}
\begin{equation}
\hat{A}_t(h_t; \psi) = \sum_{l=0}^T (\gamma \lambda)^l \delta_{t+l},
 \label{pg_adv}
\end{equation}
where $\pi'(a_t|h'_t; \phi')$ is the behavior policy for sampling trajectories, which does not participate in gradient updates. $\pi(a_t|h_t; \phi)$ is the target policy for optimization. $\frac{\pi_t}{\pi_t'}$ is the importance sampling ratio which is used to correct the distribution errors caused by the difference between the behavior and target policies. Besides, we introduce $c_{h} \mathcal{H}(\pi_t) $ as a regularization term to further encourage exploration during training, where $\mathcal{H}(\pi_t)$ denotes the entropy of the policy and $c_{h}$ is a coefficient. However, the off-policy method is known for being unstable and hard to coverage. To address this issue, the clip function, ${\rm clip}_{1-\epsilon}^{1+\epsilon}$, is used to limit the value of the importance sampling ratio by removing the incentive for moving the ratio outside of the interval $[1-\epsilon, 1+\epsilon]$, thus it can prevent very large policy updates and stabilize the training. To balance the trade-off between variance and bias of the training objective, we utilize the generalized advantage estimator (GAE) \cite{schulman2016high}, $\hat{A}_t$, as given by Eq. (\ref{pg_adv}), where $\lambda \in [0, 1]$ is used to control the trade-off between bias and variance. {\color{black} GAE can dramatically reduce the variance of the objective while keeping a tolerable bias level. }

Algorithm \ref{DRACM} summarizes the training process of the DRACM. Each training loop consists of the sampling process and the target policy updating process. In the sampling process, we firstly synchronize the parameters of the behavior and target networks (include policy network and encoder network), and then sample a set of trajectories from the environment using the behavior encoder and policy networks. The advantage estimator, $\hat{A}_t$, can then be obtained based on the sampled trajectories. Next, in the target policy updating process, we conduct training of $m$ loops to update the parameters of the encoder network, policy network, and critic network via mini-batch stochastic gradient descent with \textit{Adam} \cite{kingma2014adam}. After training, the target policy and encoder networks can be deployed to the end device for making online migration decisions by neural network inference, which has a linear time complexity of $O(n)$, where $n$ is the length of the history. In the next subsection, we discuss how to implement the DRACM in the emerging MEC system. 


\subsection{The DRACM empowered MEC framework}
\label{sec::dracm_system_archi}

\begin{figure}[t]
    \centering
    \includegraphics[width=3.3in]{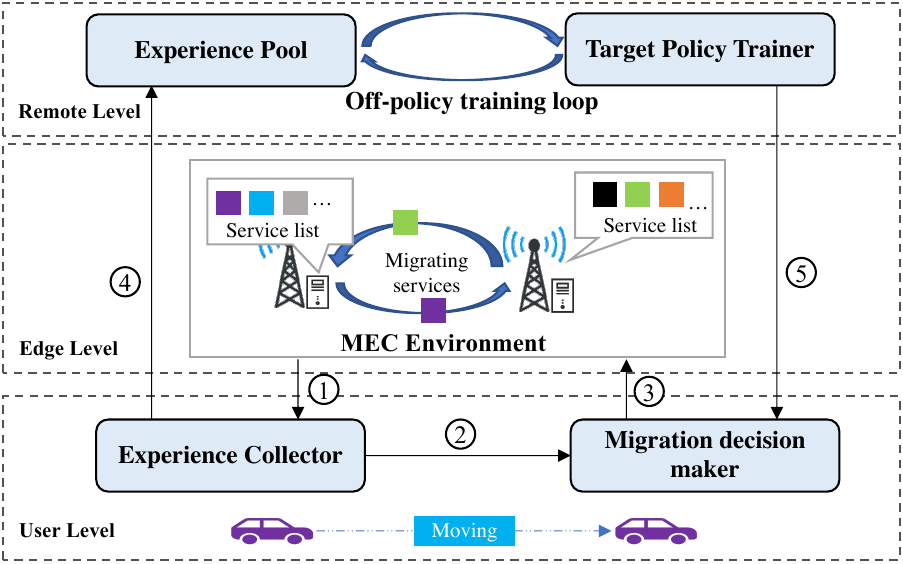}
    \centering\caption{The framework of DRACM empowered MEC system. The data flows in this framework are: \textcircled{\small{1}} the observation $o_t$ and reward $r_t$ from the MEC environment, \textcircled{\small{2}} the history $H_t = \{o_0,a_0, ..., a_{t-1}, o_t\}$ for migration decision-making, \textcircled{\small{3}} the migration action, $a_t$, made by the behavior policy, \textcircled{\small{4}} the collected trajectories uploaded to the experience pool, \textcircled{\small{5}} the parameters of the trained target policy and encoder networks for service migration. }
    \label{mec_archi_dracm}
\end{figure}

The emerging MEC system defined by ETSI consists of three levels: user level, edge level, and remote level \cite{sabella2019developing}. The user level includes various mobile devices such as smartphones and vehicles. The edge level consists of multiple edge servers where each server provides services for processing tasks that are offloaded by mobile users. The edge servers are connected through backhaul links, thus the service can be migrated among them. The remote level includes data centers with large storage and computing capacity. Fig. \ref{mec_archi_dracm} shows the overall framework of integrating the DRACM into the three-level MEC system. Four key components (\textit{experience collector}, \textit{migration decision maker}, \textit{experience pool}, and \textit{target policy trainer}) of the DRACM are deployed at the user and remote level:

\begin{itemize}
  \item At the user level, the \textit{experience collector} is responsible of collecting the information of observations and rewards from the MEC environment (Step \textcircled{\small{1}}). It sends the history $H_t = \{o_0,a_0, ..., a_{t-1}, o_t\}$ to the \textit{migration decision maker} for online decision-making (Step \textcircled{\small{2}}), and the collected trajectories to the \textit{experience pool} for the target policy training (Step \textcircled{\small{4}}). The \textit{migration decision maker} includes behavior policy and encoder networks. It downloads parameters from the \textit{target policy trainer} as the initial values of the behavior policy and encoder networks (Step \textcircled{\small{5}}), and decides the migration actions based on the observed history (Step \textcircled{\small{3}}). 
  \item At the remote level, the \textit{experience pool} stores the sampled trajectories from mobile users. The \textit{target policy trainer} is in charge of training the target policy based on the sampled trajectories. 
\end{itemize}

\begin{figure*}[t]
    \centering
    \begin{subfigure}[b]{0.30\textwidth}
    \includegraphics[width=2.1in]{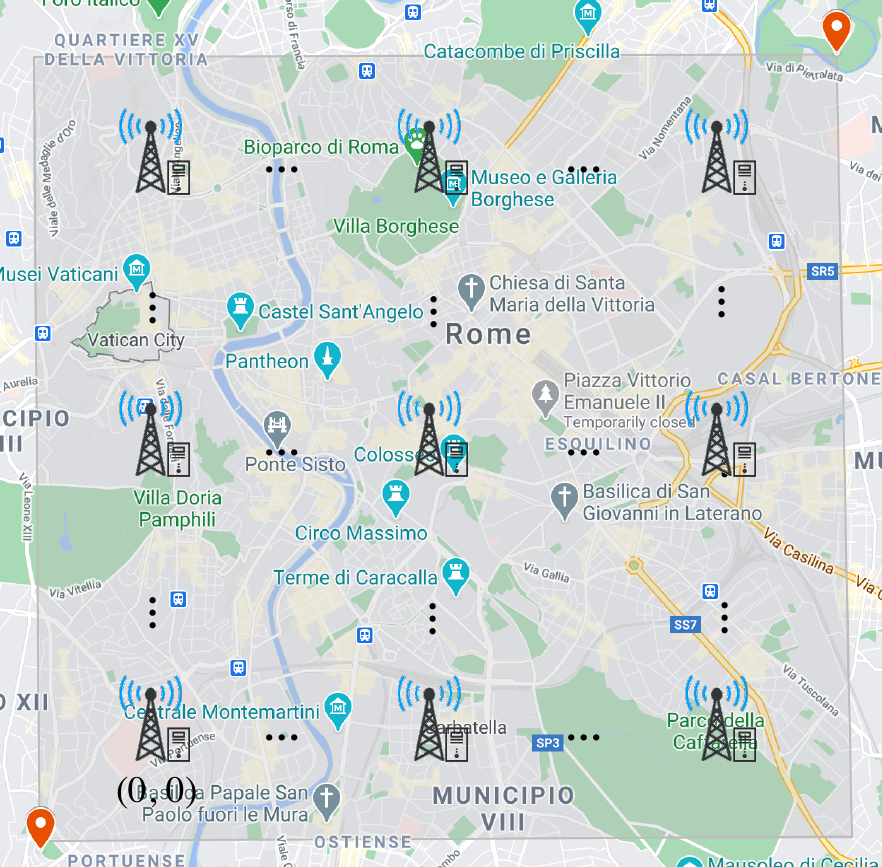}
    \end{subfigure}
    \begin{subfigure}[b]{0.30\textwidth}
    \includegraphics[width=2.0in]{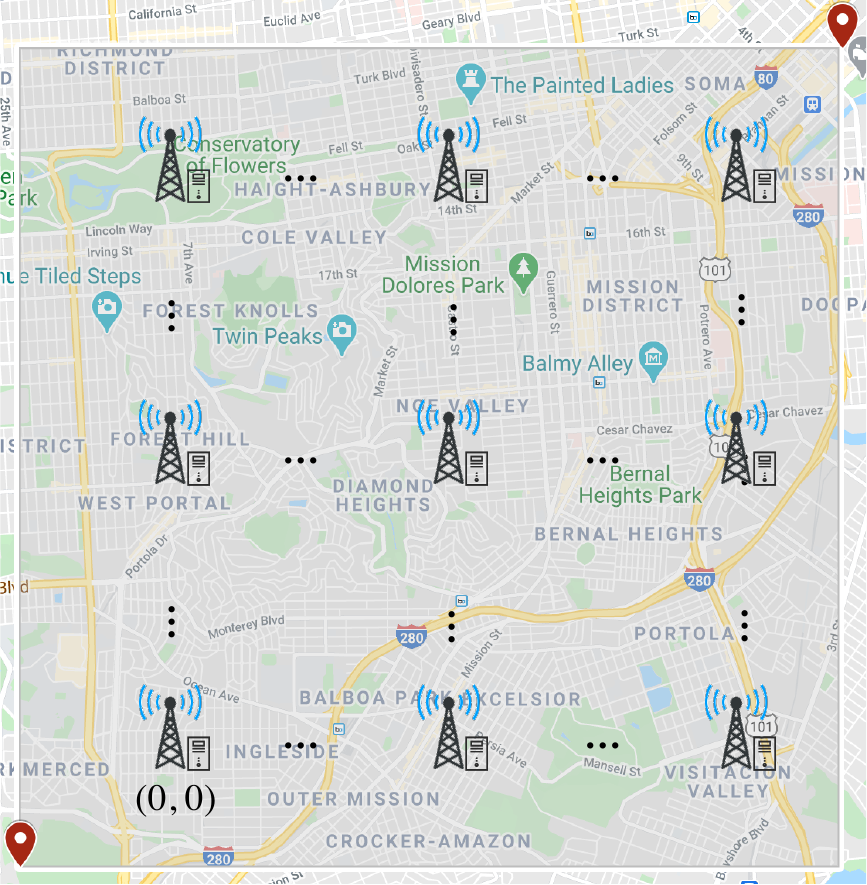}
    \end{subfigure}
    \centering\caption{The central areas of Rome, Italy (8 km $\times$ 8 km area bounded by the coordinate pairs [41.856, 12.442] and [41.928, 12.5387]) and San Francisco (8 km $\times$ 8 km area bounded by the coordinates pairs [37.709, -122.483] and [37.781, -122.391]).}
    \label{map}
\end{figure*}

According to Algorithm \ref{DRACM}, the \textit{target policy trainer} conducts multiple training loops with mini-batch gradient updates based on the collected trajectories in the \textit{experience pool}. Note that the training can be offline without directly interacting with the MEC environment. After training, the \textit{target policy trainer} sends the updated parameters of policy and encoder networks to mobile users for the next-round of sampling process.

\section{Experiments}
\label{sec::experiments}

In this section, we present the comprehensive evaluation results of the DRACM in detail. Our experiments demonstrate that: 1) the DRACM has a stable and efficient training process; 2) the DRACM can autonomously adapt to different MEC scenarios including various user's task arriving rates, applications' processing densities, and coefficients of migration delay. We firstly introduce the experiment settings based on a real-world MEC environment. Next, we present the baseline algorithms for comparison. Finally, we evaluate the performance of the DRACM and baseline algorithms in different MEC scenarios. 

\subsection{Experiment settings}
We evaluate the DRACM with two real-world mobility traces of cabs in Rome, Italy \cite{roma-taxi-20140717} and San Francisco, USA \cite{piorkowski2009crawdad}. Specifically, we focus our analysis to the central parts of Rome and San Franscisco, as shown in Fig. \ref{map}. We consider that 64 MEC servers are deployed in each area, where each MEC server covers a 1 km $\times$ 1 km grid with a computation capacity $f = 128$ GHz (i.e., four 16-core servers with 2 GHz for each core). According to \cite{narayanan2020first}, the upload rate of real-world commercial 5G networks is generally less than 60 Mbps. Therefore, in our environment, the upload rate $\rho_t$ in each grid is set as 60, 48, 36, 24, and 12 Mbps from a proximal end to a distal end. The hop distances between two MEC servers are calculated by Manhattan distance. The location of an MEC server is represented by a 2-D vector $(i, j)$ with respect to a reference location at $(0, 0)$. To calculate the propagation latency, we set the bandwidth of backhaul network, $\eta_t$, as 500 Mbps \cite{ma2019efficient} and the coefficient of backhaul delay, $\lambda_{\rm bh}$, as 0.02 s/hop \cite{yuan2020joint}. The migration delay varies with various service types and network conditions, e.g., the migration delay of Busybox (a type of service) ranges from 2.4 to 3.3 seconds \cite{ma2019efficient} with different bachkhaul bandwidths. Following some related work on MEC \cite{ouyang2019adaptive,wang2019delay,ma2019efficient}, we assume the coefficient of migration delay is uniformly distributed in $[1.0, 3.0]$ s/hop during our training. 

\begin{table}[t]
  \small
  \begin{center}
    \caption{Parameters of the Simulated Environment.}
    \begin{tabular}{>{\centering\arraybackslash}p{2.0in} | >{\centering\arraybackslash}p{1.0in}} 
      \hline
      \textbf{Parameter} & \textbf{Value}  \\ 
      \hline
      Computation capacity of an MEC server, $f$ & 128 GHz \\
      \hline
      Upload rate of wireless network, $\rho_t$ & \{60, 48, 36, 24, 12\} Mbps \\
      \hline
      Bandwidth of backhaul network, $\eta_t$ & 500 Mbps \\
      \hline
      Coefficient of backhaul delay, $\lambda_{\rm bh}$ & 0.02 s/hop  \\
      \hline
      Coefficient of migration delay, $m_{t}^{c}$ & $U[1.0, 3.0]$ s/hop \\
      \hline
      Data size of each offloaded task & $U[0.05, 5]$ MB \\
      \hline
      Processing density of an offloaded task, $\kappa$ & $U[200, 10000]$ cycles/bit \\
      \hline
      User's task arriving rate $\lambda_{p}^{u}$ & 2 tasks/slot \\
      \hline
      MEC server's task arriving rate $\lambda_{p}^{s}$ & $U[5, 20]$ tasks/slot \\
      \hline
    \end{tabular}
    \label{parameter_environment}
  \end{center}
\end{table}

\begin{table}[t]
  \small
  \begin{center}
  \caption{Hyperparameters of the DRACM.}
    \begin{tabular*}{0.49\textwidth}{c | c || c | c} 
      \hline
      \textbf{Hyperparameter} & \textbf{Value} & \textbf{Hyperparameter} & \textbf{Value} \\ 
      \hline
      LSTM Hidd. Units & 256 & Embedding Dim. $d_e$ & 2\\
      \hline
      Actor Layer Type & Dense &Actor Hidd. Units & 128 \\
      \hline
      Critic Layer Type & Dense &Critic Hidd. Units & 128 \\
      \hline
      Learning Rate & 0.0005 &  Optimizer & Adam \\
      \hline
      Discount $\lambda$ & 0.95 &  Discount $\gamma$ & 0.99 \\
      \hline
      Coefficient $c_{h}$ & 0.01 & Clipping Value $\epsilon$ & 0.2  \\
      \hline
    \end{tabular*}
    \label{hyperparameters_setting}
  \end{center}
\end{table}

At each time slot, the tasks arriving at a mobile user and those arriving at an MEC server are sampled from Poisson distributions with rates $\lambda_{p}^{u}$ and $\lambda_{p}^{s}$, respectively. In our experiments, we show the performance of the DRACM under different task arriving rates of mobile users. According to the current works \cite{nguyen2020privacy,chen2015efficient,zhan2020mobility}, the data size of an offloaded task in real-world mobile applications often varies from 50 KB (sensor data) \cite{nguyen2020privacy} to 5 MB (image data) \cite{chen2015efficient}. Hence, we set the data size of each offloaded task uniformly distributed in $[0.05, 5]$ MB. The required CPU cycles of each task can be calculated by the product of the data size and processing density, $\kappa$, which is uniformly distributed in $[200, 10000]$ cycles/bit, covering a wide range of tasks from low to high computation complexity \cite{kwak2015dream}. We summarize the parameter settings of our simulation environment in Table \ref{parameter_environment}. 
\begin{figure}[t]
    \centering
    \includegraphics[width=2.85in]{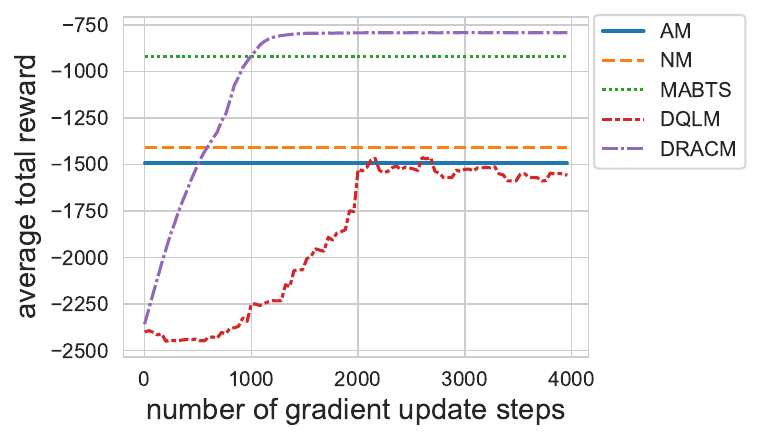}
    \caption{Average total reward of the DRACM and baseline algorithms with the mobility traces of Rome.}
    \label{training_rome}
\end{figure}

\begin{figure}[t]
    \centering
    \includegraphics[width=2.85in]{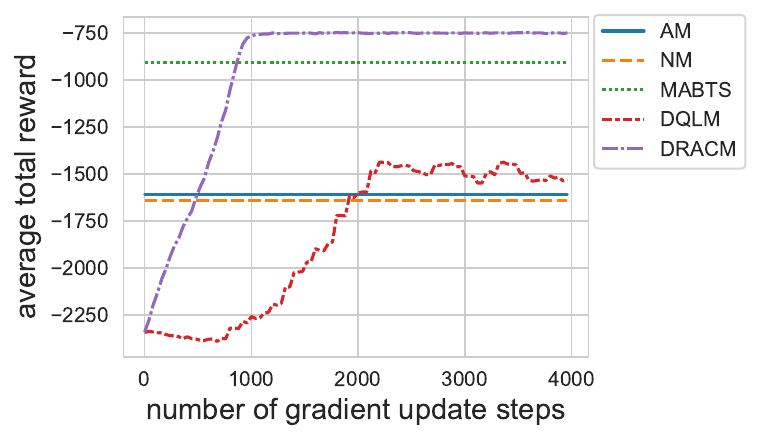}
    \caption{Average total reward of the DRACM and baseline algorithms with the mobility traces of San Francisco.}
    \label{training_san}
\end{figure}

\begin{figure}[t]
    \centering
    \includegraphics[width=2.5in]{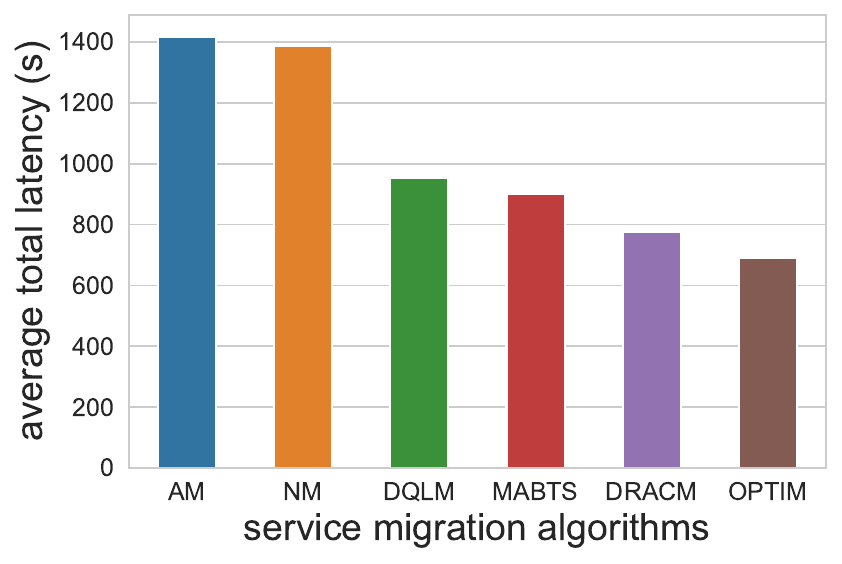}
    \caption{Average total latency (s) of service migration over the time horizon (250 minutes) on the testing dataset from mobility traces of Rome.}
    \label{testing_results_rome}
\end{figure}

\begin{figure}[t]
    \centering
    \includegraphics[width=2.5in]{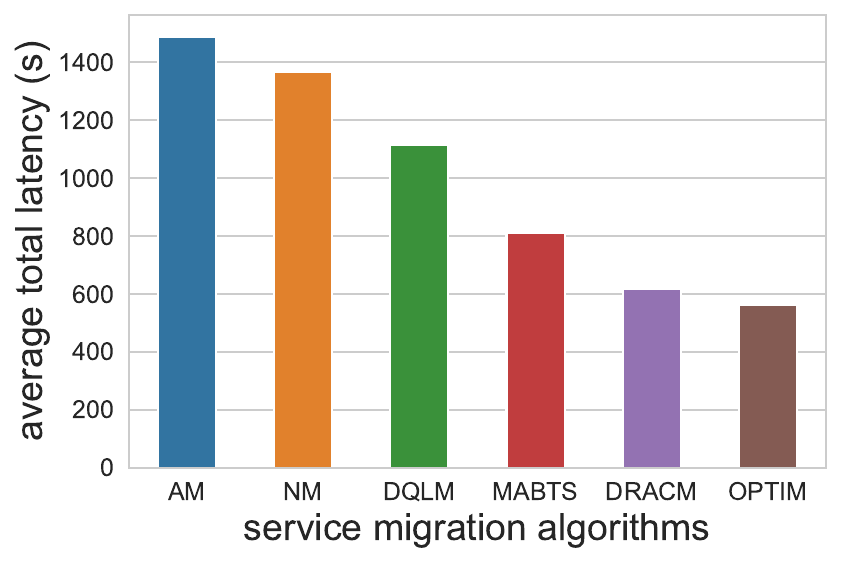}
    \caption{Average total latency (s) of service migration over the time horizon (250 minutes) on the testing dataset from mobility traces of San Francisco.}
    \label{testing_results_san}
\end{figure}

\subsection{Baseline algorithms}
We compare the performance of the DRACM to that of five baseline algorithms:
\begin{itemize}
    \item \textbf{Always migrate (AM): } A mobile user always selects the nearest MEC server to migrate at each time slot. 
    \item \textbf{Never migrate (NM): } The service is placed on an MEC server and never migrate during the time horizon. 
    \item \textbf{Multi-armed Bandit with Thompson Sampling (MABTS)}: Some exiting works \cite{sun2017emm,ouyang2019adaptive} solve the service migration problem based on MAB. According to the work \cite{ouyang2019adaptive}, MABTS uses a diagonal Gaussian distribution to approximate the posterior of the cost for each arm and applies Thompson sampling to handle the trade-off between exploring and exploiting. 
    \item \textbf{DQL-based migrate (DQLM): }Some recent works \cite{wang2019delay,wu2020mobility,yuan2020joint} adapt DQL to tackle the service migration problem. For a fair comparison, we use similar neural network structure as DRACM to approximate the action-value function for DQLM, but use the objective function of the DQL method as the training target. Moreover, we use $\epsilon$-greedy to control the exploring-exploiting trade-off as the above works do. 
    \item \textbf{Optimal migrate (OPTIM): }Assuming the user mobility trace and the complete system-level information over the time horizon are known ahead, the service migration problem can be transformed to the shortest-path problem \cite{ouyang2018follow,wang2019delay}, which can be solved by the Dijkstra algorithm. 
\end{itemize}
The NM, AM, MABTS, and DQLM algorithms can run online, while the OPTIM is an offline algorithm which defines the performance upper-bound of service migration algorithms.

\subsection{Evaluation of the DRACM and baseline algorithms}

\begin{figure}[t]
    \centering
    \includegraphics[width=2.85in]{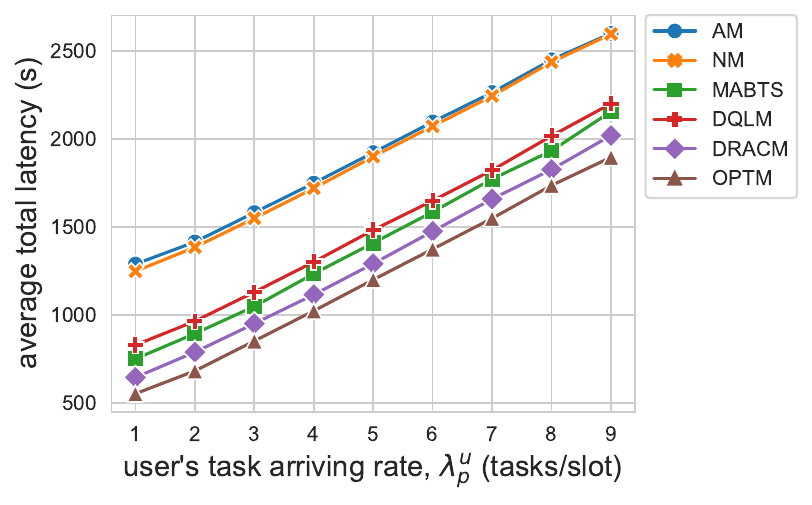}
    \caption{Average total latency (s) of service migration over the time horizon (250 minutes) with different task arriving rates of users (mobility traces of Rome).}
    \label{testing_results_arriving_rate_rome}
\end{figure}

\begin{figure}[t]
    \centering
    \includegraphics[width=2.85in]{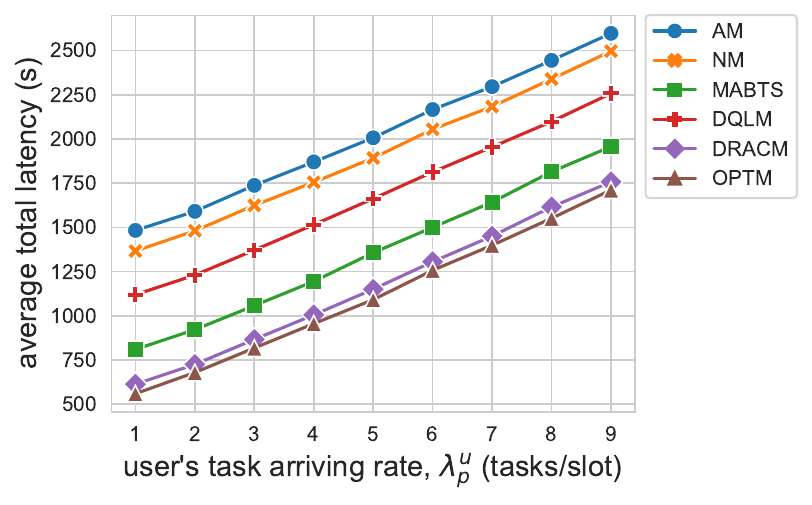}
    \caption{Average total latency (s) of service migration over the time horizon (250 minutes) with different task arriving rates of users (mobility traces of San Francisco).}
    \label{testing_results_arriving_rate_san}
\end{figure}

\begin{figure}[t]
    \centering
    \includegraphics[width=2.85in]{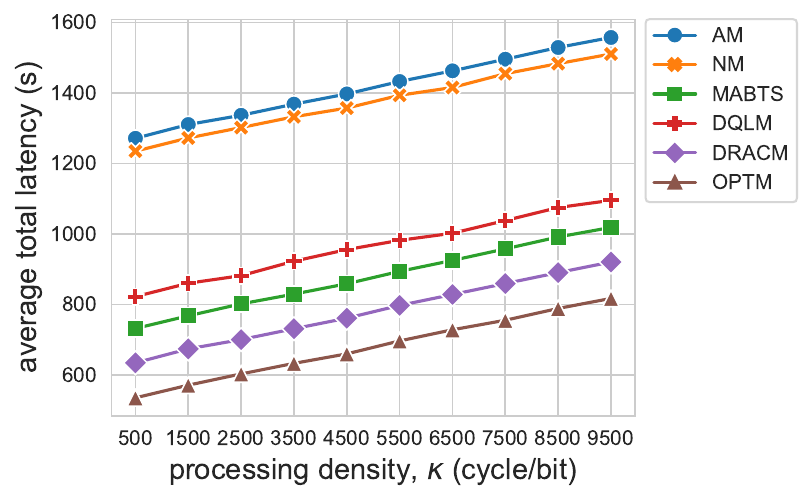}
    \caption{Average total latency (s) of service migration over the time horizon (250 minutes) with different processing densities (mobility traces of Rome).}
    \label{testing_results_request_pattern_rome}
\end{figure}

\begin{figure}[t]
    \centering
    \includegraphics[width=2.85in]{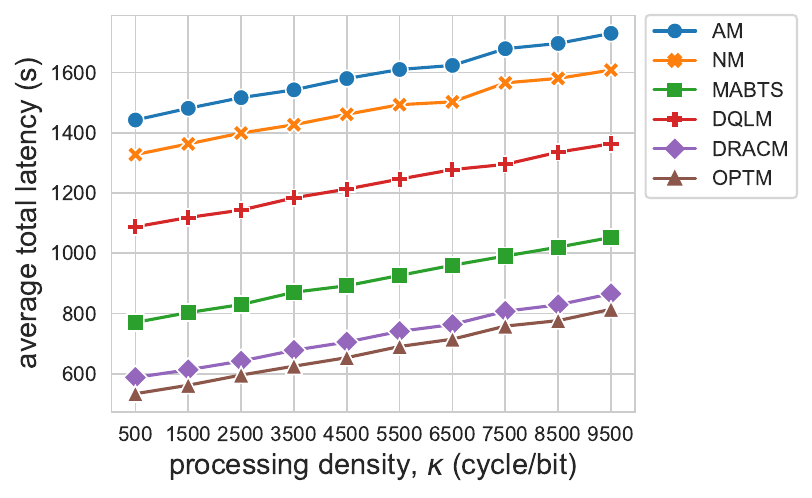}
    \caption{Average total latency (s) of service migration over the time horizon (250 minutes) with different processing densities (mobility traces of San Francisco).}
    \label{testing_results_request_pattern_san}
\end{figure}

We first evaluate the training performance of the DRACM and DQLM on two different mobility trace datasets \cite{roma-taxi-20140717,piorkowski2009crawdad}. Each training dataset includes 100 randomly picked mobility traces, where each trace has 100 time slots of three-minute length each. Table \ref{hyperparameters_setting} lists the hyperparameters in training. The neural network structure of the DQLM is similar to the DRACM with the same encoder network. The difference is that, rather than using the actor-critic structure, the DQLM is based on the Q-network that includes a fully connected layer with 128 hidden units to approximate the action-value function and chooses the action with the largest action-value at each time step. We train the DQLM and DRACM with the same learning rate, mini-batch size, and number of gradient update steps.

Figs. \ref{training_rome} and \ref{training_san} show the training results of DRACM and DQLM on mobility traces of Rome and San Francisco, respectively. The other baseline algorithms do not involve the training process for neural networks, thus we show their final performance. The network parameters of both DRACM and DQLM are initialized by random values, thus they randomly select actions to explore the environment and achieve the worst results compared to other baseline algorithms before training. However, the DRACM quickly surpasses NM and AM after 12 epochs and keeps growing on both mobility traces. After 25 training epochs, the average total reward of the DRACM remains stable, which shows the excellent convergence property of the DRACM. Besides, the final stable results of the DRACM on both mobility traces beat all baseline algorithms. 

To evaluate the generalization ability of the DRACM, we test the trained target policy on testing datasets of both mobility traces, where each test dataset includes 30 randomly picked mobility traces that were not included in the training dataset. Figs. \ref{testing_results_rome} and \ref{testing_results_san} present the results of the average total latency of DRACM and baseline algorithms on Rome and San Francisco mobility traces, respectively. We found the DRACM achieves the best performance compared to online baseline algorithms on both mobility traces. Specifically, Fig. \ref{testing_results_rome} shows that the DRACM outperforms the DQLM and MABTS by 18\% and 13\%, respectively. Fig. \ref{testing_results_san} indicates that the DRACM surpasses the DQLM and MABTS by 44\% and 23\%, respectively. Furthermore, the DRACM achieves near-optimal results within 12\% of the optimum on both mobility traces.

We then test the DRACM and baseline algorithms with different task arriving rates of users on both mobility traces. As shown in Figs. \ref{testing_results_arriving_rate_rome} and \ref{testing_results_arriving_rate_san}, the average total latencies of all evaluated algorithms increase with the rise of user's task arriving rate, since the average number of offloaded tasks increases at each time slot. The evaluation results show that the DRACM adapts well among different task arriving rates of users, where it outperforms the DQLM and MABTS by up to 24\% and 45\%, respectively. Moreover, in all cases, the results of DRACM are close to the optimal values.  

Next, we investigate the performance of the DRACM with different processing densities. For a real-world mobile application, the higher is the processing density, the more computation power is required for processing the application. Figs. \ref{testing_results_request_pattern_rome} and \ref{testing_results_request_pattern_san} depict the average total latency of DRACM on Rome mobility traces and San Francisco mobility traces, respectively. We find that the DRACM adapts well to the change of processing density on both mobility traces, where it outperforms all online baselines.

\begin{figure}[t]
    \centering
    \includegraphics[width=2.85in]{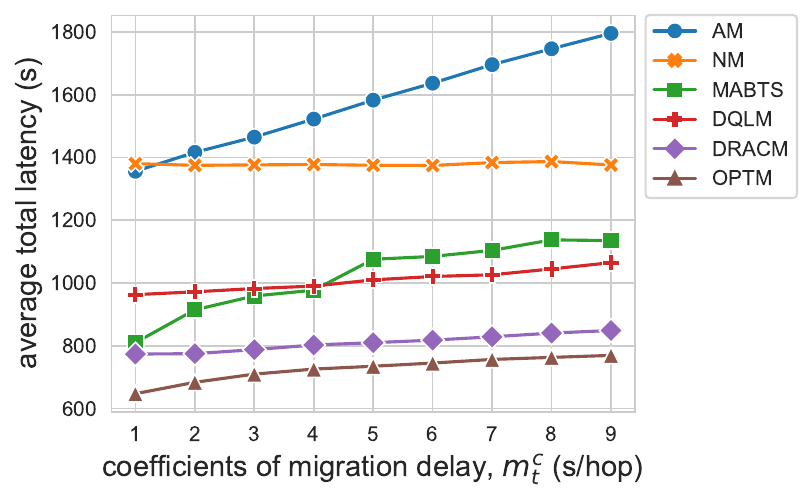}
    \caption{Average total latency (s) of service migration over the time horizon (250 minutes) with different coefficients of migration delay (mobility traces of Rome).}
    \label{testing_results_migration_coefficient_rome}
\end{figure}

\begin{figure}[t]
    \centering
    \includegraphics[width=2.85in]{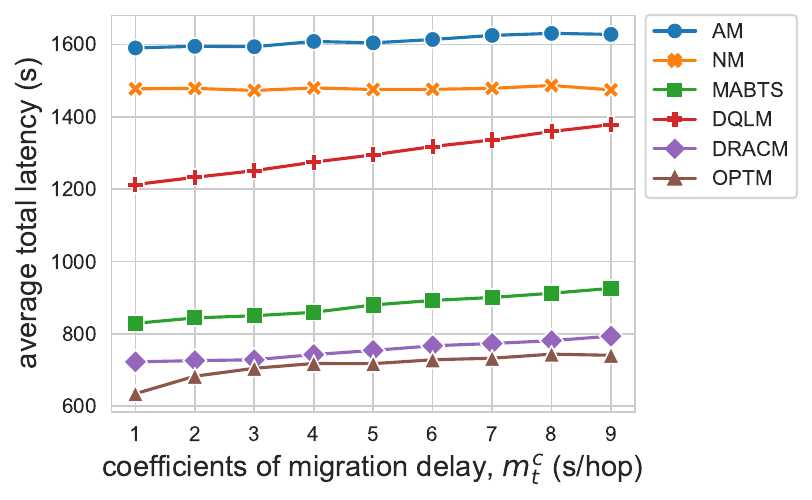}
    \caption{Average total latency (s) of service migration over the time horizon (250 minutes) with different coefficients of migration delay (mobility traces of San Francisco).}
    \label{testing_results_migration_coefficient_san}
\end{figure}

Migration delay is another important factor that influences the overall latency. To investigate the impact of the migration delay, we evaluate the DRACM and baseline algorithms on the testing datasets with different coefficients of migration delay. Intuitively, when the migration delay is high, a mobile user may not choose to frequently migrate services. As shown in Figs. \ref{testing_results_migration_coefficient_rome} and \ref{testing_results_migration_coefficient_san}, the NM algorithm keeps the similar performance in all cases while the performance of other algorithms drops with the increase of $m^{t}_{c}$. This is because that the NM does not involve the migration process and thus has no migration delay. In Fig. \ref{testing_results_migration_coefficient_rome}, we find the MABTS suffers serious performance degradation as $m_t^c$ increases. When the $m^{t}_{c}$ is low (e.g., $m^{t}_{c}=1.0$), the MABTS achieves similar results as the DRACM. However, when $m^{t}_{c} > 4$, the performance of MABTS becomes even worse than the DQLM. Compared to RL-based methods like the DQLM and DRACM, MABTS is ``short-sighted'' since it only considers the one-step reward rather than explicitly optimizes the total reward over the entire time horizon. Overall, the DRACM autonomously learns to adapt among the scenarios with different migration delays, which achieves the best performance compared to the online baselines (with up to 25\% improvement over the MABTS and up to 42\% improvement over the DQLM), and obtains near-optimal results in our experiments. 

 The DRACM method has many advantages: 1) the learning-based nature of the DRACM makes it flexible among different scenarios with few human expertise; 2) the user-centric design is scalable for the increasing number of mobile users, where each mobile user makes effective online migration decisions based on the incomplete system information; 3) the tailored off-policy training objective improves both performance and stability of the training process; 4) the design of online decision-making and offline policy training makes the DRACM more practical in real-world MEC systems. Beyond the scope of service migration, the framework of the DRACM has the potential to be applied to solve more decision-making problems in MEC systems such as task offloading and resource allocation \cite{mao2017survey}.

\section{Related Work}
\label{sec::related_work}
Service migration in MEC has attracted intensive research interests in recent years. Rejiba et al. \cite{rejiba2019survey} published a comprehensive survey on mobility-induced service migration in fog, edge, and related computing paradigms. We roughly classify the related work into centralized control approach (the central cloud or MEC servers make service migration decisions for all mobile users) and decentralized control approach (each mobile user makes its own migration decisions). 

\textbf{Centralized control approach:} plenty of works focused on making centralized migration decisions (i.e., the migration decisions are made by ether central cloud or edge servers) based on the complete system-level information to minimize the total cost. Ouyang et al. \cite{ouyang2018follow} converted the service migration problem as an online queue stability control problem and applied Lyapunov optimization to solve it. {\color{black} Ning et al. \cite{ning2020distributed} formulate the service migration problem by jointly considering the constraints of server storage capability and service execution latency. They utilize Lyapunov optimization and distributed Markov approximation to enable dynamic service placement. Liu et al. \cite{liu2020distributed} propose a multi-agent RL based method for the service migration where agents represent the controllers of MEC servers. } Xu et al. \cite{xu2020path} formulated the service migration problem as a multi-objective optimization framework and proposed a method to achieve a weak Pareto optimal solution. Wang et al. \cite{wang2019dynamic} formulated the service migration problem as a finite-state MDP and proposed an approximation of the underlying state space. They solve the finite-state MDP by using a modified policy-iteration algorithm. Other recent works tackled the service migration problem based on RL. Wang et al. \cite{wang2019delay} proposed a Q-learning based micro-service migration algorithm in mobile edge computing. Wu et al. \cite{wu2020mobility} considered jointly optimizing the task offloading and service migration, and proposed a Q-learning based method combing the predicted user mobility. These works considered the case where the decision-making agent knows the complete system-level information. However, in a practical MEC system, collecting complete system-level information can be difficult and time-consuming. Moreover, the centralized control approach may suffer from the scalability issue when facing a rapidly increasing number of mobile users. 

\textbf{Decentralized control approach:} some studies proposed to make migration decisions by the user side based on incomplete system-level information. Ouyang et al. \cite{ouyang2019adaptive} formulated the service migration problem as an MAB and proposed a Thompson-sampling based algorithm that explores the dynamic MEC environment to make adaptive service migration decisions. Sun et al. \cite{sun2018learning} proposed an MAB based service placement framework for vehicle cloud computing, which can enable the vehicle to learn to select effective neighboring vehicles for its service. Sun et al. \cite{sun2017emm} developed a user-centric service migration framework using MAB and Lyapunov optimization to minimize the latency with constraints of energy consumption. These methods simplify the system dynamics by modeling with MAB, which ignores the inherently large state space and complex transitions among states in a real-world MEC system. Distinguished from the above works, our method models the service migration problem as a POMDP that has a continuous state space and models complex transitions between states. Moreover, our method is model-free and adaptive to different scenarios, which can learn to make online service migration decisions with minimal expert knowledge. More recently, Yuan et al. \cite{yuan2020joint} investigated the joint service migration and mobility optimization problem for vehicular edge computing. They modeled the MEC environment as a POMDP and proposed a multi-agent DRL method based on independent Q-learning to learn the policy. However, using Q-learning based method to solve the environment with complex dynamics and continuous state space can be unstable and inefficient. Our evaluation results show that our method can achieve stabler training and better results than the DQL-based method.

\section{Conclusion}
\label{sec::conclusion}
In this paper, we proposed the DRACM, a new method for solving the service migration problem in MEC given incomplete system-level information. Our method is completely model-free and can learn to make online migration decisions through end-to-end RL training with minimal human expertise. Specifically, the service migration problem in MEC is modeled as a POMDP. To solve the POMDP, we designed an encoder network that combines an LSTM and an embedding matrix to effectively extract hidden information from sampled histories. Besides, we proposed a tailored off-policy actor-critic algorithm with a clipped surrogate objective to improve the training performance. We demonstrated the implementation of the DRACM in the emerging MEC framework, where migration decisions can be made online from the user side and the training for the policy can be offline without directly interacting with the environment. We evaluated the DRACM and four online baseline algorithms with real-world datasets and demonstrated that the DRACM consistently outperforms the online baselines and achieves near-optimal results on a diverse set of scenarios.

\bibliographystyle{IEEEtran} 
\bibliography{pomdp-service-migration}

%








\end{document}